\documentclass[traditabstract,letter]{aa}
\usepackage{txfonts}
\usepackage{graphicx}
\usepackage{natbib}
\newcommand{\kms}{\mbox{km~s$^{-1}$}}
\newcommand{\mols}{\mbox{molec.~s$^{-1}$}}

\begin{document}

%   \title{Detection of ethylene glycol and formamide in comets C/2012~F6 (Lemmon)
%and C/2013~R1 (Lovejoy)
   \title{Complex organic molecules in comets C/2012~F6 (Lemmon)
and C/2013~R1 (Lovejoy): detection of ethylene glycol and formamide 
\thanks{Based on observations carried out with the IRAM 30-m telescope.
IRAM is supported by INSU/CNRS (France), MPG (Germany) and IGN (Spain).}
}

   \author{N. Biver\inst{1}
        \and D. Bockel\'ee-Morvan\inst{1}
        \and V. Debout\inst{1}
        \and J. Crovisier\inst{1}
        \and J. Boissier\inst{2}
        \and D.C. Lis\inst{3,4}
        \and N. Dello Russo\inst{5}
        \and R. Moreno\inst{1}
        \and P. Colom\inst{1}
        \and G. Paubert\inst{6}
        \and R. Vervack\inst{5}
        \and H.A. Weaver\inst{5}}

   \institute{LESIA, Observatoire de Paris, CNRS, UPMC, Universit\'e 
        Paris-Diderot, 5 place Jules Janssen, F-92195 Meudon, France
   \and IRAM, 300, rue de la Piscine, F-38406 Saint Martin d'H\`eres, France
   \and California Institute of Technology, Cahill Center for
         Astronomy and Astrophysics 301-17, Pasadena, CA~91125, USA
   \and Sorbonne Universit\'{e}s, UPMC, %Universit\'{e} Pierre et Marie Curie Paris 6,
        CNRS, Observatoire de Paris, LERMA, F-75014 Paris, France
   \and JHU/APL, Laurel, Maryland, USA
   \and IRAM, Avd. Divina Pastora, 7, 18012 Granada, Spain
 }

   \titlerunning{Glycol and formamide in comets C/2012~F6 and C/2013~R1}
   \authorrunning{Biver et al.}
   \date{May 26, 2014}

   \abstract{A spectral survey in the 1~mm wavelength range was
undertaken in the long-period comets C/2012~F6
(Lemmon) and C/2013~R1 (Lovejoy) using the 30~m telescope of the
Institut de radioastronomie millim\'etrique (IRAM) in April and
November-December 2013. We report the detection of
ethylene glycol (CH$_2$OH)$_2$ ({\it aGg'} conformer) and formamide (NH$_2$CHO) 
in the two comets. The abundances  relative to water of ethylene
glycol and formamide  are 0.2--0.3\% and 0.02\% in the two
comets, similar to the values measured in comet C/1995 O1
(Hale-Bopp). We also report the detection of HCOOH and CH$_3$CHO
in comet C/2013~R1 (Lovejoy), and a search for other complex
species (methyl formate, glycolaldehyde).

%Its abundance is compared to that of other detected organic
%molecules (CH$_3$OH, H$_2$CO, HCOOH)

}
% {}{}{}{}

\maketitle

%----------------------------------------------------------------------
\section{Introduction}

Comets are the most pristine remnants of the formation of the
solar system 4.6 billion years ago. Investigating the
composition of cometary nuclei ices provides clues to the physical
conditions and chemical processes at play in the primitive solar
nebula. Comets may also have played a role in the delivery of
water  and organic material to the early Earth \citep{Har11}.

%Establishing their detailed composition is important for
%understanding how comets may have seeded the early Earth with
%prebiotic molecules and contributed to the origin of life.

The recent years have seen significant improvement in the sensitivity
and spectral coverage of millimetre receivers. The EMIR
receivers \citep{Car12} at the Institut de radioastronomie millim\'etrique
(IRAM) are equipped with a fast Fourier-transform
spectrometer that offers a wide frequency coverage at a high
spectral resolution (0.2~MHz). The combination enables sensitive
spectral surveys of cometary atmospheres and searches for
complex molecules through their multiple rotational lines in the 1~mm band.

We report here the detection of ethylene glycol
(CH$_2$OH)$_2$ and formamide (NH$_2$CHO) in comets C/2012~F6
(Lemmon) (hereafter referred to as Lemmon) and C/2013~R1
(Lovejoy) (hereafter Lovejoy), for the first time since their discovery in
comet C/1995 O1 (Hale-Bopp) \citep{Cro04a,Boc00}. We also present
the detection of CH$_3$CHO and HCOOH in comet Lovejoy. Comet
Lemmon is a long-period comet \citep[initial orbital period of
9800 yr;]%, inclination of 83\deg;]
[]{MPEC2014E18}, which reached
perihelion at 0.731 AU on 24 March 2013. Comet Lovejoy passed
perihelion on 22 December 2013 at 0.812 AU from the Sun and 
also is a high-inclination long-period comet originating from the
Oort Cloud \citep[initial orbital period of 7000 yr;] %, inclination of 64\deg;]
[]{MPEC2014E18}. Comets Lemmon and Lovejoy became
unexpectedly bright naked-eye comets, reaching visual magnitudes of
4.5 and 4.8, and displayed high water production
rates near perihelion of 1$\times10^{30}$ and $1.5\times10^{29}$
\mols \citep[][Crovisier et al., personal communication]{Com14}.

\section{Observations}
Data on comet Lemmon were acquired with the IRAM 30-m telescope on March
14--18 and April 6--8, 2013, when favourable weather conditions enabled 
observations. Although one of the most productive
comets of the past years, comet Lemmon was relatively far from 
Earth near perihelion ($\Delta$ = 1.5 AU).

Comet Lovejoy was discovered on 7 September 2013 \citep{Lov13},
which was two and a half months before its optimal observing conditions
around perigee at $\Delta$ = 0.397 AU on 19 November. It was
observed during three periods: 8--12 November, 27 November--1 December, 
and 9--16 December, 2013, under average to good weather
conditions. Additional observations obtained on November 13 and 16,
which focused on the search for phosphine, were reported by
\citet{Agu14}. During this period, the water production rate of
comet Lovejoy increased from $\sim5$ to $\sim11\times 10^{28}$
\mols~according to Nan\c{c}ay OH observations (Crovisier et al.,
personal communication). The best sensitivity was reached at the
end of November when the comet was still close to Earth and
more active than two weeks earlier.

For the two comets, we surveyed most of the 1~mm band from 210 to
272~GHz with five different double-sideband tunings. Each
tuning covers $2\times8$~GHz in two linear
polarizations, with the upper 8~GHz sideband separated by 8~GHz
from the lower sideband. The 166--170~GHz range was also
observed in both comets. The comets were tracked with the latest
available orbital elements, pointing was checked once every 1-2
hours, and residual pointing offsets were estimated from coarse maps
of the strongest cometary lines. The HCN $J$(3--2) line
at 265.886~GHz and methanol lines present in all tuning setups
were observed regularly to monitor the activity and track the 
location of the peak
of gas emissions in the coma. Table~\ref{tabobsefe} provides the
geometric circumstances of the observations together with
reference coma and outgassing parameters derived from
observations.

\begin{table*}
\caption[]{Observing circumstances and reference
parameters.}\label{tabobsefe}\vspace{-0.2cm} \centering
\begin{tabular}{lccccccc}
\hline\hline\noalign{\smallskip}
UT date  & $<r_{h}>$  & $<\Delta>$   & $v_{exp}$  & $T_{kin}$ & $Q_{\rm H_2O}$ & $Q_{\rm HCN}$ & $Q_{\rm CH_3OH}$  \\
$($aaaa/mm/dd.d--dd.d) & (AU)  & (AU)  & (\kms)  & (K) & (\mols) & (\mols) & (\mols) \\
\hline\noalign{\smallskip}
\multicolumn{8}{l}{Comet C/2012~F6 (Lemmon):} \\
2013/04/06.4--08.6   & 0.78 & 1.60 &  1.00 & 100 & $\sim9\times10^{29}$$^a$ &  $12.7\pm0.9\times10^{26}$ & $14.6\pm1.1\times10^{27}$ \\
\multicolumn{8}{l}{Comet C/2013~R1 (Lovejoy):} \\
2013/11/08.2--12.4   & 1.13 & 0.47 & 0.76 &  55 &  $0.5\times10^{29}$$^b$ & $7.4\pm0.1\times10^{25}$ &  $1.4\pm0.1\times10^{27}$ \\
2013/11/27.5--31.6   & 0.92 & 0.47  & 0.90 &  65 & $0.8\times10^{29}$$^b$ & $12.7\pm0.1\times10^{25}$ &  $2.2\pm0.2\times10^{27}$ \\
2013/12/09.5--16.6   & 0.83 & 0.71  & 0.93 &  80 &  $1.1\times10^{29}$$^b$ & $18.0\pm0.1\times10^{25}$ &  $2.9\pm0.2\times10^{27}$ \\
\hline
\end{tabular}\vspace{-0.2cm}
\tablefoot{$^{(a)}$ From \citet{Com14} and in agreement with $Q_{\rm CH_3OH}$,
and the values of $Q_{\rm H_2O}$ and $Q_{\rm CH_3OH}$/$Q_{\rm
H_2O}$ measured by \citet{Pag14}. $^{(b)}$ From contemporaneous
Nan\c{c}ay OH observations (Crovisier et al., personal com.).}
\end{table*}

\begin{figure}[h]
\sidecaption
\includegraphics*[width=7.0cm,angle=0]{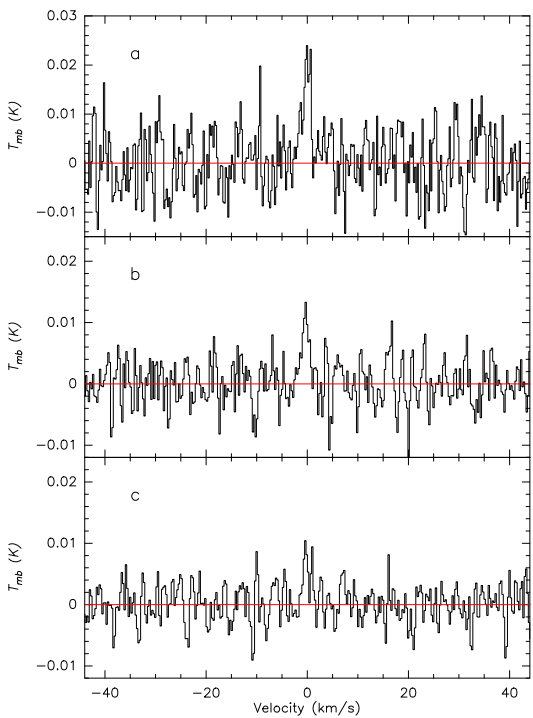}
\caption{Weighted average of the 9--13 strongest {\it aGg'} (CH$_2$OH)$_2$
lines observed in the 213--272 GHz range (online Table~\ref{tabglycol}):
(a) C/2012 F6 (Lemmon), 6--8 April 2013, (b) C/2013~R1 (Lovejoy),
27 Nov.--1 Dec. 2013; (c) Lovejoy, 9--16 Dec. 2013. The vertical
scale is the main beam brightness temperature and the
horizontal scale is the Doppler velocity in the comet frame.
\vspace{-0.5cm}} \label{figglycol}
\end{figure}

\begin{figure}[h]
\includegraphics*[width=6.3cm,angle=270]{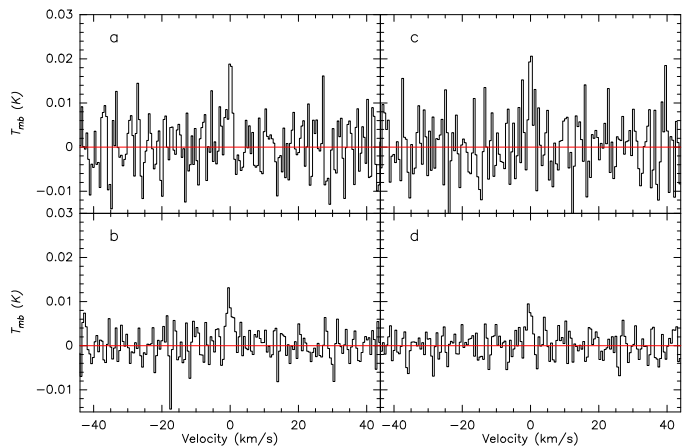}
\caption{Weighted average of the 5--10 strongest NH$_2$CHO $J$(11--10) and
$J$(12--11) lines (online Table~\ref{tabformamide}). (a) C/2012~F6 (Lemmon),
6--8 April 2013; (b) C/2013~R1 (Lovejoy), 8--12 Nov. 2013; (c)
C/2013~R1 (Lovejoy), 27 Nov.--1 Dec. 2013; (d) C/2013~R1
(Lovejoy), 9--16 Dec. 2013. Scales as in Fig.\ref{figglycol}.}
\label{fignh2cho}
\end{figure}

\section{Data analysis}
\label{sec:reduction}

Spectra were corrected for the main beam efficiency (0.55
to 0.44 in the 210--272 GHz range).
Calibration was regularly checked on reference line sources (W3OH,
W51D and IRC+10216) and the main beam efficiency was checked
several times per day on planets (e.g., Mars, Uranus). Losses due
to bad focus (up to 20\% during some daytime observations) and
elevation dependence of the antenna gain were corrected. The beam
size varies between 11.7\arcsec~and 9.1\arcsec~from 210 to
272~GHz, which corresponds to $\approx$12\,000 and 3\,000--6\,000~km 
at the distances of comets Lemmon and Lovejoy.

%, to avoid being biased towards a single line, especially to
%derive the production rate.

  Several individual lines of ``classical'' molecules such as HCN, HNC,
CH$_3$CN, CH$_3$OH, H$_2$CO, CS, and H$_2$S (and CO in Lovejoy) were
clearly detected \citep{Biv13,Agu14}. To search for
molecules with weak signatures, we averaged the multiple lines
present in the covered frequency range, focusing on the
strongest lines with similar expected intensities. As far as
possible, only lines for which the noise level
was similar were considered. Intensities were computed following
the models of \citet{Biv99,Biv00,Biv06,Biv11}, using the coma
temperatures and velocities given in Table~\ref{tabobsefe}.  For
comet Lemmon, the 240--272~GHz range was much less noisy, while
most of the 210--272~GHz range could be used for Lovejoy. Several
molecules previously identified in a few comets were detected using
this method, e.g., HNCO, OCS, HC$_3$N, SO, and SO$_2$ in comet
Lemmon, and HNCO and HC$_3$N in comet Lovejoy 
\citep[e.g.,][and {\it in preparation}]{Biv13}.

Here, we focus on the detection of ethylene glycol
(CH$_2$OH)$_2$ in its lowest energy conformer {\it aGg'}, and of
formamide (NH$_2$CHO) in the two comets. Spectra showing the
average of 5 to 13 lines are shown in
Figs.~\ref{figglycol}--\ref{fignh2cho}. Table~\ref{tabqprod}
lists average line intensities and inferred production rates. 
The detailed information on the lines considered in the analysis
is given in online Tables~\ref{tabglycol}--\ref{tabformamide}. 
Excitation processes considered in
calculating the production rates of these two molecules are
collisional excitation (with neutrals and electrons)
and spontaneous decay. Infrared pumping by
Sun radiation is not considered because band strengths are not
available in the literature. Generally, infrared pumping tends 
to decrease the effect of spontaneous radiative decay, but for the present 
observations the difference with full LTE is very small.  

 The gas kinetic temperature
(Table~\ref{tabobsefe}) was estimated from the series of methanol
rotational lines around 251~GHz: 18 lines sampling energy levels
with $E_{\rm up}$ between 60 and 240~K, and 28 lines with $E_{\rm
up}$ in the range 60--360~K were detected in comets Lovejoy and
Lemmon. The retrieved kinetic temperatures 
do not differ significantly (by more than 10\%), whether derived from other 
series of methanol lines observed between 165 and 305 GHz, from 
methanol lines at offset positions ($\approx$7\arcsec), 
or from other molecules (H$_2$S, CS). Values derived from 
251~GHz and 165~GHz methanol lines in comet Lovejoy agree within 3~K.
The gas expansion velocity
(Table~\ref{tabobsefe}) is deduced from the line width of the
strongest emission lines (e.g., HCN $J$(3--2), CS $J$(5--4) and
some strong CH$_3$OH lines).

Section~\ref{other} presents the detection of HCOOH and CH$_3$CHO and 
searches for other CHO-bearing molecules.

\begin{table*}
\caption[]{Average line intensities and production rates of ethylene glycol,
formamide, acetaldehyde, and formic acid.}\label{tabqprod}\vspace{-0.2cm} \centering
\setlength{\tabcolsep}{0.1mm}
\begin{tabular}{lllccccccccc}
\hline\hline\noalign{\smallskip}
Molecule & Freq.   & Lines$^a$ & \multicolumn{2}{c}{C/2012 F6 (Lemmon)} & \phantom{0} & & & \multicolumn{2}{c}{C/2013 R1 (Lovejoy)} && \\
 \cline{4-5}  \cline{7-12}  \\[-0.2cm]
 & range  &   & \multicolumn{2}{c}{6--8 Apr.} && \multicolumn{2}{c}{8--12 Nov.} & \multicolumn{2}{c}{27--31 Nov.} & \multicolumn{2}{c}{9--16 Dec.} \\
& & & $I$$^b$  & $Q$  & & $I$$^b$ & $Q$ & $I$$^b$ & $Q$ & $I$$^b$ & $Q$ \\
&(GHz)         &       & (mK~\kms) &
($10^{26}$s$^{-1}$)\phantom{0}& &(mK~\kms) & ($10^{26}$s$^{-1}$) &
(mK~\kms) &
($10^{26}$s$^{-1}$) & (mK~\kms) & ($10^{26}$s$^{-1}$)  \\
\hline\noalign{\smallskip}
(CH$_2$OH)$_2$\phantom{0}& 213--241\phantom{0} & 7(6$^{c}$) &-- & -- &   & $<42$ & $<5.3$ & $18\pm6$ & $2.5\pm0.8$  & $18\pm4$ &$3.8\pm0.8$ \\
(CH$_2$OH)$_2$ & 241--270 & 9        &  $39\pm5$ & $22\pm3$ & &--& --& -- & --& -- & --\\
(CH$_2$OH)$_2$ & 245--268 & 6     &  --   & --  & &$<21$ & $<3.9$  & $23\pm4$ & $4.0\pm0.7$  & $13\pm4$ & $3.0\pm0.9$ \\
NH$_2$CHO & 223--240 & 5 & --       & --       & &$<32$    & $<6.4$      & $38\pm12$& $3.2\pm1.0$ & $17\pm5$ & $3.1\pm0.9$ \\
NH$_2$CHO & 243--261 & 5  & $24\pm7$ & $15\pm4$ && $34\pm8$ & $2.2\pm0.5$ & $21\pm4$ & $1.7\pm0.3$ & $16\pm5$ & $2.4\pm0.7$ \\
NH$_2$CHO & 263--267 & 2  & $<54$    & $<28$ &   & $<42$    & $<2.6$      & $22\pm8$ & $1.7\pm0.6$ & $<24$    & $<3.4$ \\
CH$_3$CHO & 242--271 & 26 & 9$\pm$4 & $<7$ &  & -- & -- & -- & -- & -- & -- \\
CH$_3$CHO & 211--268 & 45(39$^{c}$) & -- & -- & & $<9$ & $<0.7$ & 9$\pm$2 & 0.8$\pm$0.2 & 7$\pm$2& 1.1$\pm$0.3  \\
HCOOH     & 241--268 &  6          & $<24$ & $<6.0$ &  & -- & -- & -- & -- & -- & --   \\
HCOOH     & 215--271 & 20(16$^{c}$) & -- & -- & & $12\pm5$ & $<0.6$ & 25$\pm$3 & 1.3$\pm$0.2 & 10$\pm$2& 1.0$\pm$0.2  \\
 \hline
\end{tabular}\vspace{-0.2cm}
\tablefoot{$^{(a)}$ Number of lines -- details are provided in online Tables~\ref{tabglycol}--\ref{tabformamide}.
%Line frequencies were taken from the Cologne Database for Molecular Spectroscopy \citep{CDMS}
$^{(b)}$ Weighted average of line integrated intensities $\int{T_{mb}dv}$. 
$^{(c)}$ For the 27--31 Nov. period.}
%$^{(c)}$ The strongest selected lines are $J_{K,J,1}$--$(J-1)_{K,(J-1),0}$ and
%$J_{K,J,0}-(J-1)_{K,(J-1),1}$ doublets ($K$ = 0 and 1) for $J$ =
%22 to 29 and the doublets at 230577.7 MHz (both comets), 253617.7,
%and 269942.1~MHz (comet Lemmon). $^{(d)}$ The 222348.9~MHz line
%was not observed on 27--31 Nov.. $^{(e)}$
%$J_{Ka,Kc}$--$(J-1)_{Ka,Kc-1}$ transitions with $Ka \leq 2$ and
%$J$ = 11. $^{(f)}$ with $J$ = 12. $^{(g)}$ with $J$ = 13. $^{(h)}$
%For the 27--31 Nov. period.}
\end{table*}

\subsection{Production rates of ethylene glycol and formamide}
The estimated production rates of ethylene glycol are given in
Table~\ref{tabobsefe}. The predicted intensities of the ethylene
glycol lines are very close to those expected for thermal
equilibrium (the difference is lower than 1\%), so that the
approximations used for the excitation of this molecule probably do not
affect our estimates of the production rates significantly. The
photodissociation rate at 1 AU was assumed to be $\beta_0 =
2.0\times10^{-5}$ s$^{-1}$ \citep{Cro04a}. Multiplying it by 2
increases the derived production rates by $\approx20$\% for Lemmon and
$\approx10$\% for Lovejoy.

The width and asymmetry of the ethylene glycol line detected in comet
Lemmon (Fig.~\ref{figglycol}a) is comparable to that of the other
species. The derived production rate is indicative of an abundance
relative to water of 0.24\%.

Ethylene glycol was detected in comet Lovejoy during the second
(27--31 Nov.) and third (9--16 Dec.) observing periods with
average productions rates of $3.4\pm1.1\times10^{26}$ and
$3.4\pm0.6\times10^{26}$~\mols, yielding abundances
relative to water of $0.40\pm0.13$\% and $0.30\pm0.05$\%.
The marginal difference between the two periods might
be due to, e.g., an underestimate of the gas temperature
$T_{\rm kin}$ when the comet was closer to Earth at the end of
November. Using $T_{\rm kin}$ = 80 K instead of 65 K, the derived
abundance for late November is 0.36$\pm$0.11\%, and the
differences between the production rates derived from the low- and
high-frequency groups of lines (Table~\ref{tabqprod}) are also
reduced. On the other hand, we did not find any measurable
variation ($>$ 10\%) of $T_{\rm kin}$ in the coma, yielding
an uncertainty lower than 5\% on the production rate of 
ethylene glycol.

For the first observing period, we derive an upper limit
of 0.62\%, which explains the non-detection. 

The derived production rate of formamide strongly depends on the
assumed photodissociation rate, which is presumably high
\citep[$\beta_0=7\times10^{-4}$ s$^{-1}$ according to][]{Jac76a,Jac76b}.
In this case, most molecules lie in the
collision-dominated coma region, so that radiative processes do
not play an important role in the rotational excitation of the
molecule. Following \citet{Boc00}, we
investigated how the assumed photodissociation rate affects the
derived production rate for comet Lovejoy and its evolution with time. 
Indeed, comet
Lovejoy was observed over a range of heliocentric and geocentric
distances (Table~\ref{tabobsefe}), which makes the evolution of the
signal sensitive to the lifetime of formamide. Abundances relative
to water for the three periods are ($7.2\pm5.0$; $5.4\pm1.9$;
$7.3\pm2.5$)$\times10^{-4}$, ($3.4\pm2.0$; $2.2\pm0.6$;
$2.0\pm0.6$)$\times10^{-4}$, and ($2.8\pm1.8$; $1.7\pm0.5$;
$1.3\pm0.4$)$\times10^{-4}$ for $\beta_0 = 7.0, 1.0$, and
$0.1\times10^{-4}$~s$^{-1}$, respectively. For
$\beta_0=1.0\times10^{-5}$~s$^{-1}$, the abundance of formamide
decreases with time, while for $\beta_0=7.0\times10^{-4}$ s$^{-1}$
the abundance increases from period 2 to 3. This means that $\beta_0$ is
more likely around $1.0\times10^{-4}$ s$^{-1}$, which implies an
abundance of formamide relative to water in Lovejoy of 0.021\%.
The detection of formamide in comet Lemmon is marginal, but this
value of $\beta_0$ yields a NH$_2$CHO/H$_2$O ratio of
$0.016\pm0.005$\%, which is close to that found in comet Lovejoy
and similar to the value measured in comet Hale-Bopp
\citep{Boc00}. Production rates given in Table~\ref{tabqprod} are
for $\beta_0$ = $1.0\times10^{-4}$ s$^{-1}$. 
A 10\% variation of $T_{\rm kin}$ would only modify  
the production rate of formamide in either comet by 3\%.

\subsection{Abundances of other complex molecules}\label{other} 
Using the method outlined in
Sect.~\ref{sec:reduction}, we searched for a number of CHO-bearing
molecules in the spectra of comets Lemmon and Lovejoy. Production
rates, or upper limits, were determined assuming the local
thermodynamical equilibrium, and the corresponding abundances relative to
water are displayed in Table~\ref{tababund}. Formic acid (HCOOH)
is detected in comet Lovejoy (more than 6 lines with a S/N $>$3),
but not in comet Lemmon. Acetaldehyde (CH$_3$CHO) is detected at
the 4$\sigma$ level in comet Lovejoy, both at the end of
November (average of 39 lines) and in December
(45 lines) (Table~\ref{tabglycol}). CH$_3$CHO was previously 
detected only in
comet Hale-Bopp before \citep{Cro04b}. Methyl formate (CH$_3$OCHO)
and glycolaldehyde (CH$_2$OHCHO) are not detected, but significant
upper limits are obtained. Methyl formate and formic acid were
detected in comet Hale-Bopp \citep{Boc00,Cro04b}. We also include
in Table~\ref{tababund} the abundances of HNCO and H$_2$CO
(Biver et al. {\it in preparation}).

\section{Discussion}

Chemical diversity is observed in the population of Oort Cloud
comets, with abundances varying by up to a factor of ten for
several simple species such as CO, CH$_3$OH, H$_2$S, CS, and
hydrocarbons \citep[e.g.,][]{Boc11}. The origin of this diversity
is unclear and might reflect comet formation at different places
and times in the early solar system. The abundances of ethylene 
glycol and formamide in comets Lemmon and
Lovejoy are remarkably similar to the values measured in comet
Hale-Bopp (Table~\ref{tababund}). Conversely, HNCO and CO
are depleted by a factor of 4--5 in both comets with respect to
Hale-Bopp, whereas HCOOH is depleted in comet Lemmon and not in
Lovejoy, and acetaldehyde is enhanced in Lovejoy.

The comparison of cometary abundances with those measured around
protostars can shed light on the processes responsible for the
formation of complex molecules in protoplanetary disks. It is important
to decipher the role of grain-surface chemistry and radiation
processing versus gas-phase chemistry \citep{Walsh2014}. As
discussed by \citet{Cro04a} and \citet{Walsh2014}, an
important issue is the now confirmed high abundance
of ethylene glycol in comets, which is at least 5--6 times
higher than that of the chemically related species CH$_2$OHCHO
(glycolaldehyde) (Table~\ref{tababund}). By contrast,
both molecules are observed in similar abundances in the hot core
Sgr B2(N) \citep{Hollis2002}, and tentatively in the Class 0 solar-type
binary protostar IRAS 16293--2422, where ethylene glycol relative
to glycolaldehyde is 0.3--0.5 \citep{Jor2012}. In IRAS
16293--2422, the glycolaldehyde lines have their origin in the
warm (200--300 K) gas close to binary components \citep{Jor2012}.
Similarly, ethylene glycol, together with other complex molecules
is found to originate from a warm region of radius $\sim$ 60 AU in
the Class 0 source NGC 1333-IRAS2a. Hence, this supports a
scenario where these two molecules are released into the gas phase
by the sublimation of grain mantles. From laboratory experiments,
\citet{Oberg2009} showed that both molecules are produced by
UV-irradiation of methanol ices mixed with CO, with the amount of ethylene
glycol relative to glycolaldehyde increasing with
decreasing CO content and being sensitive to the temperature. 
This suggests that complex molecules found in
cometary ices might have been formed from the irradiation of
CO-poor ices. We note, however, that there is no correlation between
the CO and ethylene glycol abundances in comets
(Table~\ref{tababund}).

Formamide, the simplest amide, may have been the starting
point for prebiotic synthesis \citep{Sal12}. 
Formamide has recently been
detected in the Class 0 source IRAS 16293--2422, with an abundance
10 times lower than the Hale-Bopp value \citep{Kahane2013}.
Several formation routes are proposed for NH$_2$CHO in
protoplanetary disks \citep{Kahane2013,Walsh2014}. Grain-surface
chemistry satisfactorily explains the NH$_2$CHO abundance relative
to water measured in cometary ices, as well as that of most
CHO-bearing species detected in comets \citep[excluding ethylene
glycol;][]{Walsh2014}.

In summary, the new measurements presented here show that the high 
abundance of ethylene glycol is most likely a specific property of comets. 
Molecules present in cometary ices could have formed by a wealth of
chemical processes, which will hopefully be better constrained in
the coming years by the combination of new observations, models,
and laboratory works.

\begin{table}
\caption[]{Abundances relative to water}\label{tababund}\vspace{-0.5cm}
\begin{center}
\begin{tabular}{lcccccccccc}
\hline\hline\noalign{\smallskip}
Molecule & \multicolumn{3}{c}{Abundance (\%)} \\
         & C/1995~O1$^a$ & C/2012~F6 & C/2013~R1 \\
& (Hale-Bopp) & (Lemmon) & (Lovejoy) \\
 \hline\noalign{\smallskip}
HCN              & 0.25    & 0.14    & 0.16  \\
CO               & 23        & 4.0$^b$    &  7.2$^c$\\
H$_2$CO          & 1.1     & 0.7$^c$     & 0.7$^c$  \\
CH$_3$OH         & 2.4     & 1.6        & 2.6  \\
HCOOH            & 0.09    & $<0.07$ & 0.12 \\
(CH$_2$OH)$_2$   & 0.25    & 0.24    & 0.35 \\
HNCO             & 0.10    & 0.025$^c$   & 0.021$^c$ \\
NH$_2$CHO        & 0.02    & 0.016   & 0.021  \\
HCOOCH$_3$       & 0.08    & $<0.16$ & $<0.20$ \\
CH$_3$CHO        & 0.025   & $<$ 0.07 & 0.10   \\
CH$_2$OHCHO      & $<0.04$ &  $<$ 0.08 &  $<$ 0.07\\
\hline
\end{tabular}
\end{center}\vspace{-0.5cm}
\tablefoot{$^{(a)}$ \citet{Boc00,Cro04a,Cro04b}. $^{(b)}$\citet{Pag14}.$^{(c)}$ Biver et al. {\it in preparation}.} \\
\end{table}

\begin{acknowledgements}
This research has been supported by the Programme national de 
planétologie de l'Institut des sciences de l'univers (INSU).
%We are grateful to the IRAM staff and to other observers for their assistance
%during the observations.
%IRAM is an international institute co-funded by the Centre national
%de la recherche scientifique (CNRS), the Max Planck Gesellschaft and
%the Instituto Geogr\'afico Nacional, Spain.

\end{acknowledgements}

%-------------------------------------------------------------------

\onltab{
\begin{table*}
\caption[]{Observed lines and production of ethylene glycol}\label{tabglycol}
\centering
\setlength{\tabcolsep}{0.5mm}
\begin{tabular}{lcclcclcclcclc}
\hline\hline
Transition  & frequency  & \multicolumn{3}{c}{Lemmon 6--8 Apr.} & \multicolumn{3}{c}{Lovejoy 8--12 Nov.} 
                                                                                 & \multicolumn{3}{c}{Lovejoy 27--31 Nov.} 
                                                                                 & \multicolumn{3}{c}{Lovejoy 9--16 Dec.} \\
            & [MHz]      & \multicolumn{2}{c}{[mK~\kms]$^a$} & $Q$  & \multicolumn{2}{c}{[mK~\kms]$^a$} & $Q$ 
                                                                                 & \multicolumn{2}{c}{[mK~\kms]$^a$} & $Q$ 
                                                                                 & \multicolumn{2}{c}{[mK~\kms]$^a$} & $Q$ \\
            &            & Model$^b$ & Obs. & [$10^{26}$s$^{-1}$]          & Model$^c$ & Obs. & [$10^{26}$s$^{-1}$] 
                                                                          & Model$^c$ & Obs. & [$10^{26}$s$^{-1}$]
                                                                          & Model$^c$ & Obs. & [$10^{26}$s$^{-1}$]  \\
\hline
$22_{1,22,1}-21_{1,21,0}$ & 213131.96 &      &     &  &      &        & &      &        & &      &        &  \\[-0.05cm]
$22_{0,22,1}-21_{0,21,0}$ & 213132.92 &      &     &  & 47.5 & \vline & & 40.5 & \vline & & 25.2 & \vline &  \\[0.1cm]
$24_{1,24,0}-23_{1,23,1}$ & 217449.99 &      &     &  &      & \vline & &      & \vline & &      & \vline &  \\[-0.05cm]
$24_{0,24,0}-23_{0,23,1}$ & 217450.27 &      &     &  & 37.0 & \vline & & 33.6 & \vline & & 22.1 & \vline &  \\[0.1cm]
$23_{1,23,1}-22_{1,22,0}$ & 222348.60 &      &     &  &      & \vline & &      & \vline & &      & \vline &  \\[-0.05cm]
$23_{0,23,1}-22_{0,22,0}$ & 222349.15 &      &     &  & 44.6 & \vline & &      & \vline & & 25.2 & \vline &  \\[0.1cm]
$25_{1,25,0}-24_{1,24,1}$ & 226643.30 &      &     &  &      &\vline~$<42$ &&&\vline~$18\pm6$&&&\vline~$18\pm4$ &\\[-0.05cm]
$25_{0,25,0}-24_{0,24,1}$ & 226643.46 &      &     &  & 34.1 &\vline&$<5.3$& 32.0 & \vline &$2.5\pm0.8$ & 21.8 & \vline & $3.8\pm0.8$ \\[0.1cm]
$24_{3,22,0}-23_{3,21,1}$ & 230577.15 &      &     &  &      & \vline & &      & \vline & &      & \vline &  \\[-0.05cm]
$22_{4,19,1}-21_{4,18,0}$ & 230578.26 &      &     &  & 36.7 & \vline & & 33.8 & \vline & & 22.7 & \vline &  \\[0.1cm]
$24_{1,24,1}-23_{1,23,0}$ & 231564.01 &      &     &  &      & \vline & &      & \vline & &      & \vline &  \\[-0.05cm]
$24_{0,24,1}-23_{0,23,0}$ & 231564.32 &      &     &  & 41.2 & \vline & & 37.5 & \vline & & 24.7 & \vline &  \\[0.1cm]
$25_{1,25,1}-24_{1,24,0}$ & 240778.13 &      &     &  &      & \vline & &      & \vline & &      & \vline &  \\[-0.05cm]
$25_{0,25,1}-24_{0,24,0}$ & 240778.30 & 39.0 &\vline && 37.9 & \vline & & 35.7 & \vline & & 24.4 & \vline &  \\[0.1cm]
$27_{1,27,0}-26_{1,26,1}$ & 245022.74 &      &\vline &&      &        & &      &        & &       &        &   \\[-0.05cm]
$27_{0,27,0}-26_{0,26,1}$ & 245022.79 & 34.7 &\vline && 27.6 & \vline & & 28.6 & \vline & &  21.3 & \vline &   \\[0.1cm]
$26_{1,26,1}-25_{1,25,0}$ & 249990.90 &      &\vline &&      & \vline & &      & \vline & &       & \vline &   \\[-0.05cm]
$26_{0,26,1}-25_{0,25,0}$ & 249991.00 & 37.4 &\vline && 34.1 & \vline & & 34.3 & \vline & &  24.7 & \vline &   \\[0.1cm]
$25_{3,23,1}-24_{3,22,0}$ & 253616.77 &      &\vline &&      & \vline & &      & \vline & &       & \vline &   \\[-0.05cm]
$24_{5,20,1}-23_{5,19,0}$ & 253618.60 & 34.8 &\vline &&      & \vline & &      & \vline & &       & \vline &   \\[0.1cm]
$28_{1,28,0}-27_{1,27,1}$ & 254208.72 &      &\vline~$39\pm5$&&&\vline~$<21$ &&&\vline~$23\pm4$ &&&\vline~$13\pm4$ &\\[-0.05cm]
$28_{0,28,0}-27_{0,27,1}$ & 254208.75 & 32.7 &\vline &$22\pm3$& 24.2 &\vline&$<3.9$ & 26.0 &\vline&$4.0\pm0.7$ & 20.1 &\vline& $3.0\pm0.9$ \\[0.1cm]
$27_{1,27,1}-26_{1,26,0}$ & 259202.27 &      &\vline & &      & \vline & &      & \vline & &      & \vline &   \\[-0.05cm]
$27_{0,27,1}-26_{0,26,0}$ & 259202.32 & 38.6 &\vline & & 29.9 & \vline & & 31.1 & \vline & & 23.1 & \vline &   \\[0.1cm]
$29_{1,29,0}-28_{1,28,1}$ & 263392.10 &      &\vline & &      & \vline & &      & \vline & &      & \vline &   \\[-0.05cm]
$29_{0,29,0}-28_{0,28,1}$ & 263392.12 & 32.9 &\vline & & 20.8 & \vline & & 23.1 & \vline & & 18.5 & \vline &   \\[0.1cm]
$28_{1,28,1}-27_{1,27,0}$ & 268412.15 &      &\vline & &      & \vline & &      & \vline & &      & \vline &   \\[-0.05cm]
$28_{0,28,1}-27_{0,27,0}$ & 268412.18 & 35.8 &\vline & & 26.2 & \vline & & 29.9 & \vline & & 21.7 & \vline &   \\[0.1cm]
$29_{2,28,0}-28_{2,27,1}$ & 269941.61 &      &\vline & &      &        & &      &        & &      &        &   \\[-0.05cm]
$29_{1,28,0}-28_{1,27,1}$ & 269942.57 & 30.1 &\vline & &      &        & &      &        & &      &        &   \\
\hline
\end{tabular}
\tablefoot{Frequencies are from the Cologne Database for Molecular Spectroscopy \citet{CDMS}\\
$^{(a)}$ Line integrated intensity $\int{T_{mb}dv}$;
$^{(b)}$ Model with $Q_{\rm glycol}=2\times10^{27}$\mols;
$^{(c)}$ Model with $Q_{\rm glycol}=5\times10^{26}$\mols} \\
\end{table*}
}

\onltab{
\begin{table*}
\caption[]{Observed lines and production rate of formamide}\label{tabformamide}
\centering
\setlength{\tabcolsep}{0.5mm}
\begin{tabular}{lc clc clc clc clc}
\hline\hline
Transition  & frequency  & \multicolumn{3}{c}{Lemmon 6--8 Apr.} & \multicolumn{3}{c}{Lovejoy 8--12 Nov.} 
                                                                                 & \multicolumn{3}{c}{Lovejoy 27--31 Nov.} 
                                                                                 & \multicolumn{3}{c}{Lovejoy 9--16 Dec.} \\
            & [MHz]      & \multicolumn{2}{c}{[mK~\kms]$^a$} & $Q$  & \multicolumn{2}{c}{[mK~\kms]$^a$} & $Q$ 
                                                                                 & \multicolumn{2}{c}{[mK~\kms]$^a$} & $Q$ 
                                                                                 & \multicolumn{2}{c}{[mK~\kms]$^a$} & $Q$ \\
            &            & Model$^a$ & Obs. & [$10^{25}$s$^{-1}$]          & Model$^a$ & Obs. & [$10^{25}$s$^{-1}$] 
                                                                          & Model$^a$ & Obs. & [$10^{25}$s$^{-1}$]
                                                                          & Model$^a$ & Obs. & [$10^{25}$s$^{-1}$]  \\
\hline
& & & & & & & & & & & & & \\
$11_{1,11}-10_{1,10}$   & 223452.51 &  & &    & 161.2 &\vline &      & 124.8 &\vline &           & 53.8 &\vline &           \\
$11_{0,11}-10_{0,10}$   & 227605.66 &  & &    & 151.7 &\vline &      &       &\vline~$38\pm12$ &  & 56.9 &\vline &           \\
$11_{2,10}-10_{2,9}$    & 232273.65 &  & &    & 134.4 &\vline~$<32$ & & 111.2 &\vline&$3.2\pm1.0$ & 51.7 &\vline~$17\pm5$ &   \\
$11_{2,9}-10_{2,8}$     & 237896.68 &  & &    & 139.9 &\vline &$<6.4$&       &       &           & 54.4 &\vline &$3.0\pm0.9$\\
$11_{1,10}-10_{1,9}$    & 239951.80 &  & &    & 164.1 &\vline &      &       &       &           & 60.8 &\vline &           \\
& & & & & & & & & & & & & \\
$12_{1,12}-11_{1,11}$   & 243521.04 & 16.7&\vline &        & 156.4&\vline &            & 124.2&\vline &            & 65.6&\vline &           \\
$12_{0,12}-11_{0,11}$   & 247390.72 & 17.4&\vline &        & 163.8&\vline &            & 129.8&\vline &            & 68.3&\vline &           \\ 
$12_{2,11}-11_{2,10}$   & 253165.79 & 15.7&\vline~$24\pm7$&& 136.7&\vline~$34\pm8$ &    & 112.6&\vline~$21\pm4$ &    & 61.9&\vline~$16\pm5$ &   \\
$12_{2,10}-11_{2,9}$    & 260189.09 & 17.1&\vline &$14\pm4$& 140.8&\vline & $2.2\pm0.5$& 116.5&\vline & $1.7\pm0.3$& 64.3&\vline&$2.4\pm0.7$ \\
$12_{1,11}-11_{1,10}$   & 261327.45 & 18.7&\vline &        & 164.0&\vline &            & 132.6&\vline &            & 71.5&\vline &           \\
& & & & & & & & & & & & & \\
$13_{1,13}-12_{1,12}$   & 263542.24 & 19.2&\vline~$<54$ &  & 156.9&\vline~$<42$ & & 127.2&\vline~$22\pm8$ &    & 69.9&\vline~$<24$ &    \\
$13_{0,13}-12_{0,12}$   & 267062.61 & 19.0&\vline &$<28$   & 165.5&\vline &$<2.6$ & 133.7&\vline & $1.7\pm0.6$ & 73.3&\vline & $<3.4$   \\
& & & & & & & & & & & & & \\
\hline
\end{tabular}
\tablefoot{$^{(a)}$ line intensities modelled with $Q_{\rm NH_2CHO}=10^{26}$\mols 
and photodissociation rates at 1AU $\beta_0=1.10^{-4}$s$^{-1}$;}
\end{table*}
}

\end{document}